\documentclass[amsmath,amssymb,prb,showpacs]{revtex4}

\usepackage{graphicx}

\begin{document}

\title{Interaction flow method for many-fermion systems}
\author{Carsten Honerkamp, Daniel Rohe, Sabine Andergassen, and Tilman Enss}
\affiliation{Max Planck Institute for Solid State Research, D-70569 Stuttgart, Germany}  
\date{March 25, 2004}

\begin{abstract} 
We propose an interaction flow scheme that sums up the perturbation expansion of many-particle systems by successively increasing the interaction strength. It combines the unbiasedness of renormalization group methods with the simplicity of straight-forward perturbation theory. Applying the scheme to fermions in one dimension and to the two-dimensional Hubbard model we find that at one-loop level and low temperatures there is ample agreement with previous one-loop renormalization group approaches. We furthermore present results for the momentum-dependence of spin, charge and pairing interactions in the two-dimensional Hubbard model.
\end{abstract}

\pacs{71.10.-w, 71.10.Fd, 71.10.Pm}

\maketitle

\section{Introduction}
The competition between different Fermi surface instabilities is a general feature of low-dimensional interacting lattice electron systems.  In one spatial dimension a number of analytical and numerical methods are available, whereas in two and higher dimensions perturbative renormalization group methods are often considered the least biased approach to study such systems at weak coupling and low energy scales. In recent years functional renormalization group methods  \cite{shankar,msbook,zanchi,halboth,tflow,kopietz,salmhofer,katanin,tsai,binz} have been devised, capable of capturing the interplay of various low-energy scattering processes within certain approximations. They provide a basis for a qualitative comparison of the strengths of potential instabilities. For those RG schemes which are derived from an \emph{exact} flow equation  \cite{wegner,polchinski,wetterich,msbook,salmhofer}, the strategy is to reduce the degree of approximation step by step in order to achieve quantitatively more accurate results.\cite{meden} However, the practical experience along these lines shows that each scheme has pros and cons. While easily applicable, the one-loop momentum shell schemes have difficulties describing ferromagnetic Stoner instabilities.\cite{tflow} Schemes with frequency cutoff obscure the analytical structure of correlation functions, and the temperature-flow scheme \cite{tflow} needs to be adapted to the specific type of the interaction. Thus, as long as an all-purpose weapon is not available, it is desirable to have a larger number of distinct approaches to tackle the problem from different angles.

Functional RG methods are extremely flexible concerning the choice of the flow parameter entering the quadratic part of the fermionic action.  The cutoff RG schemes are constructed by multiplying this quadratic part with a scale-dependent cutoff function. The  derivative of this function with respect to the scale defines momentum or frequency shells, which are integrated out in the course of the flow. The temperature-flow RG instead uses the temperature as the flow parameter, and the flow describes the change of the interactions as the temperature is lowered.

In this work we use the interaction strength as the flow parameter. It is instructive to switch on the interaction strength adiabatically and to follow the change in the correlations as the interaction is increased. This defines an {\em interaction flow} (IF). Straight-forward perturbation theory could in principle do this to any order, but casting the idea in the form of the functional RG has the compelling advantage that all one-loop corrections are summed up consistently. Therefore the IF scheme combines the unbiasedness of one-loop RG methods with a straight-forward relation to perturbation theory.

In this paper we apply the IF method to fermions in one dimension and to the two-dimensional (2D) Hubbard model. In the latter numerous instabilities may compete, and we can compare the nature of the infrared instabilities suggested by IF with results obtained from other RG approaches. Furthermore, the 2D Hubbard model is of general interest, in particular because of its potential relation to high-$T_c$ and other unconventional superconductors. However, we emphasize that the IF method is rather general and should be widely applicable to many kinds of interacting many-particle systems.

The paper is organized as follows: In Sec. \ref{formalism} we present the formalism for the IF scheme. In Sec. \ref{1dexamples} we apply it to interacting fermions in one spatial dimension. In Sec. \ref{implem} we describe the implementation in two spatial dimensions  for the Hubbard model on the square lattice and compare the results to those obtained in other one-loop RG schemes. Further more, we show new results for the detailed wave-vector dependence of spin and charge susceptibilities. In Sec. \ref{conclusions} we conclude, with remarks on the perspective of the new scheme concerning its applicability in quantum many-body physics.

\section{The Method} \label{formalism}

We start with a grand-canonical Hamiltonian of the form
\begin{equation}
H = \sum_{\vec{k}}\xi(\vec{k}) \,  c_{\vec{k},s}^\dagger  c_{\vec{k},s}
+ \frac{1}{2N}\sum_{\vec{k},\vec{k}',\vec{q} \atop s,s'} V(\vec{k},\vec{k}',\vec{k}+ \vec{q}) c_{\vec{k}+\vec{q},s}^\dagger  c_{\vec{k}'-\vec{q},s'}^\dagger c_{\vec{k}',s'}  c_{\vec{k},s} \label{ham} \, . \end{equation}
The $c_{\vec{k},s}$-operators annihilate electrons with wavevector $\vec{k}$ and spin projection $s_z=\pm 1/2$. $\mu$ is the chemical potential, $\xi(\vec{k}) = \epsilon(\vec{k}) - \mu$ the free dispersion of electrons hopping on a 2D square lattice measured with respect to the free Fermi surface, and $ V(\vec{k},\vec{k}',\vec{q}) $ defines the spin-rotationally invariant and wavevector-conserving interaction between electrons. For the Hubbard onsite-repulsion we have $V(\vec{k},\vec{k}',\vec{q}) =U>0$.\\

The interaction flow scheme is derived using the functional renormalization group formalism for one-particle irreducible (1PI) vertex functions.\cite{wetterich,salmhofer} It describes the change of  1PI vertex functions \emph{exactly} upon variation of a parameter $\ell$ in the quadratic part of the action $Q_\ell$. Writing Grassmann fields for the fermions and using a combined index $k=(i k_0 ,\vec{k})$ for Matsubara frequencies $k_0$ and wave vectors $\vec{k}$, with the Hamiltonian (\ref{ham}) the full quadratic part of the action reads
\begin{equation} Q = T \sum_{i\omega_n,\vec{k}} \bar{c}_k \, \left[ -i\omega_n + \xi(\vec{k})\right] c_k \, . \end{equation}

In the momentum-shell RG $Q_\ell$ is defined by supplementing $Q$ with a cutoff function depending on the running infrared cutoff  $\ell$. The temperature-flow scheme  \cite{tflow} uses the temperature $T$ appearing  in front of the frequency sum and in the Matsubara frequencies as flow parameter. The change of all vertex functions with $\ell$ is determined by an infinite hierarchy of equations, obtained from the $\ell$-dependent Legendre transform
\begin{equation} \Gamma_\ell (\phi , \bar{\phi}) = W_\ell  (\zeta, \bar{\zeta}) - \sum_n \sum_s \int
\frac{d^2k}{(2\pi)^2} \,  \left[
\bar{\phi}_s (\vec{k}, i \omega_n)  \zeta _s (\vec{k}, i \omega_n)  +
\phi_s (\vec{k}, i \omega_n)  \bar{\zeta} _s (\vec{k}, i \omega_n) \right]  \, ,
\label{legtraf} \end{equation}
of the generating functional for connected, non-amputated  $m$--point correlation functions $W^{(m)}$,
\begin{equation}
e^{-W_\ell (\zeta,\bar{\zeta})} = \int Dc D\bar{c} \; e^{-\mathcal{S}_\ell  (c, \bar{c},
\zeta,\bar{\zeta})} \, ,
\label{genfuncW}
\end{equation}
with $\mathcal{S} (c, \bar{c},
\zeta,\bar{\zeta})$ being the standard action used to generate correlation functions.\\

\begin{figure}
\begin{center}
\includegraphics[width=.6\textwidth]{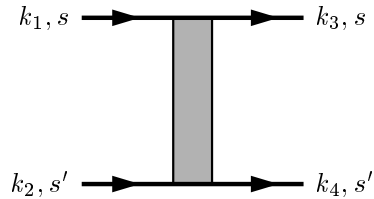}
\end{center}
\caption{The coupling function $V(\vec{k}_1,\vec{k}_2,\vec{k}_3)$. Wavevector $\vec{k}_4$
is fixed by wavevector conservation on the lattice.
 }
\label{v4}
\end{figure}

\begin{figure}
\begin{center}
\includegraphics[width=.4\textwidth]{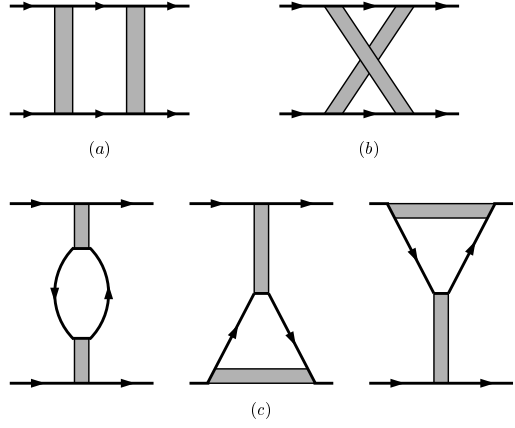}
\end{center}
\caption{The particle-particle and particle-hole diagrams contributing to the
one-loop RG equation for the interaction vertex.
 }
\label{rgdia}
\end{figure}

We condense the notation and use a combined index $p=(i \omega_n,
\vec{k},s)$, moreover $\int dp = T\sum_{n,s} \int \frac{d^2k}{(2\pi)^2}$.
The infinite hierarchy of equations is truncated in the usual way by setting the six-point function to zero for all $\ell$. The equations describing  the evolution of the self energy $\Sigma_\ell (p)$ and
the coupling function $V_\ell (p_1,p_2,p_3)$ with $p_4=p_1+p_2-p_3$ (from which the fully
antisymmetric four-point vertex can be
reconstructed \cite{salmhofer}, see also Fig.\ \ref{v4}) read
\begin{equation}
\frac{d}{d\ell} \Sigma_\ell (p) = \int dp' S_\ell(p') \,  \left[ V_\ell ( p,p',p')
- 2 V_\ell ( p,p',p) \right] \label{sigmadot} \end{equation}
and
\begin{equation} \frac{d}{d\ell} V_\ell (p_1,p_2,p_3) =    {\cal T}_{PP,\ell} + {\cal T}^d_{PH,\ell} +
{\cal T}^{cr}_{PH,\ell} \label{vdot} \end{equation}
with
\begin{eqnarray}
\lefteqn{  {\cal T}_{PP,\ell} (p_1,p_2;p_3,p_4) = }\nonumber\\
&& - \int dp \,
V_\ell ( p_1,p_2,p ) \, L(p,-p+p_1+p_2) \, V_\ell (p,-p+p_1+p_2 ,p_3) \label{PPdia}
\\[4mm]
 \lefteqn{ {\cal T}^d_{PH,\ell} (p_1,p_2;p_3,p_4) =}  \nonumber \\
&&- \int dp\,
\biggl[ -2 V_\ell ( p_1,p,p_3 ) \, L(p,p+p_1-p_3) \, V_\ell (p+p_1-p_3,p_2,p)
\nonumber
 \\ && \qquad +
V_\ell (p_1,p,p+p_1-p_3) \, L(p,p+p_1-p_3) \, V_\ell (p+p_1-p_3,
p_2,p)
 \nonumber  \\ && \qquad +
V_\ell ( p_1,p,p_3) \, L(p,p+p_1-p_3) \, V_\ell (p_2,p+p_1-p_3,p)
\biggr]\label{PHddia}
\\[4mm]
\lefteqn{  {\cal T}^{cr}_{PH,\ell}(p_1,p_2;p_3,p_4) =} \nonumber \\
&& - \int dp \,
V_\ell (p_1,p+p_2-p_3,p)  \, L(p,p+p_2-p_3) \, V_\ell (p,p_2,p_3 )\label{PHcrdia}
\end{eqnarray}
In these equations, the product of the two internal lines in the one-loop diagrams is
\begin{equation} \label{e13}
L(p,p') = S_\ell (p) W^{(2)}_{\ell} (p') + W^{(2)}_{\ell} (p) S_\ell (p')  \, \end{equation}
with the so-called single-scale propagator
\begin{equation}
S_\ell (p) = - W^{(2)}_{\ell} (p) \left[ \frac{d}{d\ell} Q_\ell
(p) \right] \, W^{(2)}_{\ell} (p) \, .
\end{equation}
The one-loop diagrams corresponding to the terms (\ref{PPdia}), (\ref{PHddia}) and
(\ref{PHcrdia})  are shown in Fig.\ \ref{rgdia}.\\

To obtain the desired interaction flow scheme we first multiply $Q$ with a scale factor $1/g$ and split it in two, yielding
\begin{equation}  Q_g = T \sum_{i \omega_n ,\vec{k}} \bar{c}_k g^{-1/2} \, \left[ -i\omega_n + \xi(\vec{k})\right] c_k g^{-1/2} \, . \end{equation}
$g$ will be the flow parameter. We can absorb the factor $1/g$ in rescaled fields $\tilde{c},\bar{\tilde{c}}$ defined as $\tilde{c} = g^{-1/2} c$. As a consequence the quartic interaction term
\begin{equation}  V^{(4)} = \frac{1}{2N}\sum_{k,k',q \atop s,s'} V(k,k',k+q) \bar{c}_{k+q,s}  \bar{c}_{k'-q,s'} c_{k',s'}  c_{k,s} \end{equation}
picks up an extra factor $g^2$ when written in terms of the new fields:
\begin{equation}  V^{(4)}_g = \frac{1}{2N}\sum_{k,k',k+q \atop s,s'} g^2 V(k,k',k+q) \bar{\tilde{c}}_{k+q,s}  \bar{\tilde{c}}_{k'-q,s'} \tilde{c}_{k',s'}  \tilde{c}_{k,s} \, . \end{equation}
Higher order interactions $V^{(2m)}$ are rescaled according to $V^{(2m)}_g =g^m V^{(2m)}$, meaning that changing the scale factor $1/g$ in $Q_g$ corresponds to changing the strength of the bare interactions. The rescaled fermions $\tilde{c}$, $\bar{\tilde{c}}$ describe a system with a bare interaction strength $g^2\,V$. In particular, we can start at an infinitely small $g$, i.e. at infinitely small bare interaction, and use the RG formalism to integrate up to the desired bare interaction strength, reached at $g=1$.  We can also stop the flow at any other value of $g$, with the functions $g \Sigma$ and  $g^2 V_g(k,k',k+q)$ being the self energy and interacting vertex function for the bare interaction $g^2 V (k,k',k+q)$. Most importantly for practical purposes, the truncation of the RG hierarchy and the subsequent neglect of self energy feedback ( see below ) does not change this property.

We call this the {\em interaction flow} (IF) scheme. Compared to straight-forward perturbation theory, the advantage of the functional RG machinery in the one-loop approximation is that it intrinsically sums up all one-loop diagrams contributing in perturbation theory. Therefore,  all kinds of ladder and bubble summations, as well as the corresponding vertex corrections are included.

\begin{figure}
\begin{center}
\includegraphics[width=.65\textwidth]{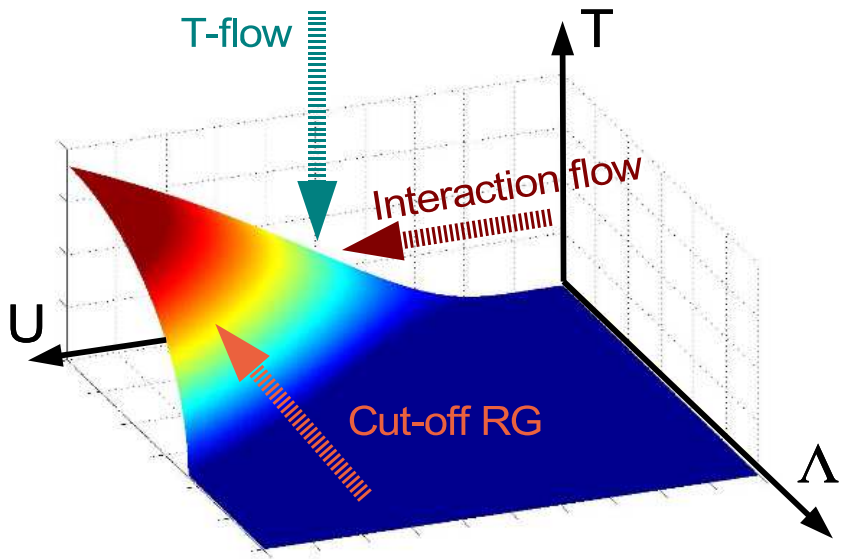}
\end{center}
\caption{Depiction of the different functional approaches in the parameter space spanned by interaction strength $g$, temperature $T$ and infrared cutoff $\Lambda$.
The surface represents the critical manifold below which the perturbation theory diverges.}
\label{RGschemes}
\end{figure}

We note that, opposed to the cutoff or $T$-flow RG schemes, singularities on the right hand side of the flow equation are not regularized by the flow parameter.
For example, the particle-particle bubble appearing in the perturbation expansion will typically diverge logarithmically when $T\to 0$. Thus, the IF scheme has to be performed at finite temperature when the individual one-loop diagrams are bounded. The strength of the interaction is increased continuously, and in the course of the interaction flow potential singularities are approached from below. 
Fig.\ \ref{RGschemes} illustrates how the various methods detect perturbative singularities from different directions in parameter space. We further remark that the IF scheme does not correspond to viewing a system on different length scales. This implies that, unlike in the cutoff RG approach, phase transitions cannot be directly related to fixed points of the flow, where the action becomes invariant upon rescaling of fields and spatial coordinates.\\

{\em Propagators:} The one-loop diagrams consist of a single-scale propagator $S_g (p)$ and full two-point Green's functions $W^{(2)}_g(p)$. When we neglect self energy corrections they are given by
\begin{equation} W^{(2)}_g (p) = g \, \left[ - i\omega_n +
\xi (\vec{k}) \right]^{-1} \end{equation}
and
\begin{equation} S_g (p) = - W^{(2)}_g (p) Q_g W^{(2)}_g (p)   =  \left[ -i \omega_n + \xi (\vec{k}) \right]^{-1} \, . \end{equation}
We have arrived at a compellingly simple prescription for the one-loop flow: we can compute the diagrams just as in perturbation theory and only need to multiply them with a scale factor  $g^{m-1}$. This shows a striking advantage of the IF scheme concerning the computational effort if self energy corrections are neglected: the one-loop corrections only depend on the flow parameter through a simple multiplicative factor $g^{m-1}$, and therefore the corresponding diagrams have to be  calculated only once in the beginning of the flow. This is in contrast to the cutoff or $T$-flow schemes, where, due to the absence of scale-invariance, the one-loop terms have to be recalculated in every RG step.

We note another crucial difference between the IF scheme and the cutoff or $T$-flow RG: in the latter the single-scale propagator receives its dominant contributions from a limited shell in the Brillouin zone, either defined by the scale-derivative of the cutoff function, or by the $T$-derivative of the Fermi function. In the course of the flow this shell contracts around the Fermi surface, such that at low scales or temperatures only quasiparticles near the Fermi surface can contribute. For weakly coupled or Fermi liquid-like systems, these are expected to be sufficiently well defined. Hence, neglecting self energy effects such as finite lifetime can be a good approximation.\cite{honedhd} In the IF scheme, however,  $S_g(p)$ is only weakly confined to a shell around the Fermi surface by the band energy in the denominator. This may lead to an overestimate of the influence of Brillouin zone regions away from the Fermi surface and should be corrected in future improvements of the method by inclusion of lifetime effects.\\

{\em Simple example:} To see how infrared singularities are treated in the IF scheme, consider a simple Cooper instability for a local attraction $V_0<0$ between electrons. This coupling constant will be subject to one-loop corrections when the scale factor $g$ is increased, starting from an infinitely small initial value.  We restrict the flow to the particle-particle channel for electrons with incoming momenta $k$ and $-k$. In this case the diagram corresponding to  ${\cal T}_{PP,\ell} (k,-k,k',-k')$ leads to a flow of the scattering vertex $V_g (k,-k,k')=V^{P}_g$ given by
\begin{eqnarray}
\frac{d}{dg} V_g^P &=& - V_g^{P} \int dp \left[ S_g(p) W^{(2)}_g(-p)+  W^{(2)}_g(p) S_g(-p) \right]  \\
&=& - V_g^{P} \,  2g \chi_{PP}(0)
\end{eqnarray}
where $\chi_{PP} (0)= \int dp \frac{1}{\omega_n^2 + \epsilon (\vec{p})^2}$ is the particle-particle bubble for zero total momentum and frequency. This equation can be integrated from 0 to $g$, yielding
\begin{equation}
 g^2\,V_g^P = \frac{g^2\,V_0}{1+ g^2 V_0 \chi_{PP} (0)} \, .
 \end{equation}
The Cooper instability occurs for $g^2 V_0 =-1/\chi_{PP} (0)$,  corresponding to the usual criterion at $g=1$. This illustrates how the instability is approached by increasing the bare coupling $g^2 V_0$. The critical bare coupling depends on temperature through the temperature dependence of the one-loop diagrams, which in the Cooper case is $\chi_{PP} (0)$. Since $\chi_{PP} (0)$ diverges logarithmically for $T \to 0 $, the critical coupling becomes infinitely small for $T \to 0$.

All other typical infrared instabilities due to summations of bubble or ladders are contained in the IF scheme in a similar way. In particular, particle-hole instabilities at small wave vectors $q$, such as ferromagnetism or Pomeranchuk Fermi surface deformations \cite{halboth}, are treated on equal footing with Cooper and large-$q$ instabilities. Thus, the IF scheme combines advantages of the temperature-flow RG with a very lucid formalism closely related to standard perturbation theory.\\

{\em Effective coupling constants in charge, spin and pairing channels:} It is useful to define certain classes of coupling functions, which drive certain susceptibilities through one-loop corrections. If any of these classes diverges, it signals a potential instability in the corresponding channel. We define effective couplings
\begin{equation} V_c(\vec{k},\vec{k}',\vec{q}) = 2 V(\vec{k},\vec{k}',\vec{k}+\vec{q}) -V(\vec{k}',\vec{k},\vec{k}+\vec{q}) \end{equation}
in the charge channel,
\begin{equation} V_s(\vec{k},\vec{k}',\vec{q}) = -V(\vec{k}',\vec{k},\vec{k}+\vec{q}) \end{equation}
in the spin channel,  and
\begin{equation} V_{\mathrm{SC}} (\vec{k},\vec{k}',\vec{q}) = V(\vec{k},-\vec{k}+\vec{q},\vec{k}') \end{equation}
in the pairing channel. The flow of these quantities can be analyzed, and their wavevector dependence contains information on the dominant tendencies of the system. For example, a $d_{x^2-y^2}$-wave Cooper instability leads to a $d_{x^2-y^2}$ dependence of  $V_{\mathrm{SC}} (\vec{k},\vec{k}',\vec{q}=0)$ on $\vec{k}$ and $\vec{k}'$, not necessarily of a separable form.

We have dropped the frequency by writing $\vec{k}$ instead of $k$, in anticipation of the following applications in which we neglect the frequency dependence of the interaction vertex.\\

{\em Coupling to external perturbations, susceptibilities:} We introduce fermionic coupling terms
to order parameters or static external fields corresponding to different symmetry breaking channels. Together with the flow of the interaction vertex we compute the one-loop
renormalizations of these coupling terms and thereby obtain information
on the growth of the corresponding fluctuations during the flow. For details about this procedure see e.g. Ref.\ \onlinecite{honsfr}. A divergence of one of these external couplings signals a
crossover to a region with a strong tendency to ordering. This could
become an actual finite temperature phase transition if allowed, e.g. when coupling in a third spatial direction is added.

The $d$-wave pairing field
$\tilde \Phi_{d\mathrm{-sc}}(\vec{k})$,
couples to the rescaled fermions at interaction strength $g^2$ and wavevector $\vec{k}$ via
\begin{equation}
\tilde\Phi_{d\mathrm{-sc}}  \,  h_{d\mathrm{-sc}}(\vec{k})  \, \left(  \bar{\tilde c}_{\vec{k},\uparrow}  \bar{\tilde c}_{-\vec{k},\downarrow} -  \bar{\tilde c}_{\vec{k},\downarrow}  \bar{\tilde c}_{-\vec{k},\uparrow} \right) \, ,
\label{phidsc}
\end{equation}
with a coupling constant $h_{d\mathrm{-sc}}(\vec{k})$. The initial
condition is taken to be a $d_{x^2-y^2}$ form factor $
h_{d\mathrm{-sc}}(\vec{k}) =\left( \cos k_x - \cos k_y \right) /
\sqrt{2}$. As for the other coupling constants defined below, the
$\vec{k}$-dependence of $h_{d\mathrm{-sc}}(\vec{k})$ will evolve through corrections of the Cooper coupling $V_{\mathrm{SC}} (\vec{k},\vec{k}',\vec{q}=0)$ in the course of the flow,  and higher harmonics of the same representation of the point group as well as a dependence on the distance to the Fermi surface will be generated.  Notice that we already wrote $\tilde\Phi_{s,d\mathrm{-sc}}$, related to the source in the IF action via $\tilde\Phi_{d\mathrm{-sc}} = g \Phi_{d\mathrm{-sc}}$. The susceptibility $\tilde\chi_{d\mathrm{-sc}}$ in this channel at interaction strength $g^2$ is obtained from the susceptibility  $\chi_{d\mathrm{-sc}}$ computed in the IF formalism as $\tilde\chi_{d\mathrm{-sc}} = \chi_{d\mathrm{-sc}}/g^2$. One can check this relation by considering the non-interacting case, where $g$ merely rescales the free fermionic Green's function.

We also define spin-density and charge-density wave fields $\vec{\tilde\Phi}^l_{s} (\vec{k},\vec{q})$ and $\vec{\tilde\Phi}^l_{c} (\vec{k},\vec{q})$ with point group symmetry $l$. They are renormalized by one-loop corrections involving the spin and charge couplings $ V_s(\vec{k},\vec{k}',\vec{q})$ and $ V_c(\vec{k},\vec{k}',\vec{q})$, respectively.
Due to spin rotation invariance it is sufficient to consider spin fields in the spin quantization direction, coupling via
\begin{equation}
\tilde\Phi^l_{z,s} (\vec{k},\vec{q}) \, h^s_{s}(\vec{k},\vec{q}) \,
\left( \bar{\tilde c}_{\vec{k}+\vec{q},s}  \tilde{c}_{\vec{k},s} - \bar{\tilde c}_{\vec{k}+\vec{q},-s} \tilde c_{\vec{k},-s} \right) \, ,
\label{phisdw}
\end{equation}
where in the $s$-wave channel the initial condition is taken to be $h^s_{s}(\vec{k},\vec{q})=1$.\\

In the 2D Hubbard model the momentum transfer $\vec{q} = \vec{Q} = (\pi,\pi)$ between created and
annihilated particles plays an important role. It corresponds to an alternating field acting on electron spins on n.n. sites, and if the corresponding coupling $h^s_{s}(\vec{k},\vec{Q})$ becomes large, this signals a strong tendency towards antiferromagnetic (AF) spin-density wave (SDW) order. We also consider couplings in the $d$-wave channels, in order to look for e.g. $d$-density wave fluctuations, coupling to particle-hole pairs in the charge channel with momentum transfer $\vec{Q}$:
\begin{equation}
\tilde\Phi^d_{c} (\vec{k},\vec{Q}) \, \sum_{s}   h^d_{c}(\vec{k})\,
\bar{\tilde c}_{\vec{k}+\vec{Q},s}   \tilde c_{\vec{k},s}\, ,
\label{phiddw}
\end{equation}
with initial condition chosen as
$ h^d_c(\vec{k}) =
\left( \cos  k_x - \cos
k_y \right) / \sqrt{2} $.
In addition to the $(\pi,\pi)$ channels, the $\vec{q}=0$-channels will also be analyzed. The isotropic $s$-wave part contains information about the compressibility of the Fermi surface, while higher $s$-wave, $d$- and $p$-wave parts are important when estimating tendencies towards Fermi surface deformation.\cite{halboth}

\section{Interacting fermions in one dimension} \label{1dexamples}

Here we briefly describe how the IF scheme reproduces some well-known perturbative results for electrons in one spatial dimension. Consider the $g$-ology model with spin-independent interactions where left- and right-moving electrons around two Fermi points $\pm k_F$ are scattered by coupling constants $g_1=V(k_F,s;-k_F,s' \to -k_F,s;k_F,s') $, $g_2=V(k_F,s;-k_F,s' \to k_F,s;-k_F,s')$ and $g_4=V(k_F,s;k_F,s' \to k_F,s;k_F,s')$. For densities sufficiently away from half filling Umklapp processes $g_3=V(k_F,s;k_F,s' \to -k_F,s;-k_F,s')$ can be neglected.  One-loop corrections in the particle-hole channel with wavevector transfer $2k_F$ and in the zero-total-momentum particle-particle channel diverge at low temperatures like $\log (W/T)$, where $W$ denotes the band width. Focusing on these contributions, and writing $s$ instead of the scale parameter $g$ in order to avoid a $g$-mess,  the one-loop IF equations read  (using $N_0= 1/(\pi v_F)$ for the density of states)
\begin{eqnarray}
\frac{dg_{1,s}}{ds} &=& - N_0 \, g_{1,s}^2 \,  2s \, \log \frac{W}{T}\label{g1eq} \\
\frac{dg_{2,s}}{ds} &=& - \frac{N_0}{2}\, g_{1,s}^2 \, 2s \, \log \frac{W}{T} \label{g2eq}
\end{eqnarray}
$g_4$ is not renormalized by logarithmic corrections.
When we now integrate form zero to the desired bare interaction strength reached at $s=1$, eq. \ref{g1eq} immediately yields (with the bare coupling $g_1 = g_{1, s=0}$)
\begin{equation} g_{1,s=1} = \frac{g_1}{1 + N_0 g_1 \log \frac{W}{T} } \, . \end{equation}
This means that at low $T \ll W$ a repulsive $g_1$ is renormalized away by one-loop contributions. The low-temperature physics is thus determined by the remaining coupling constant $g_2$ which approaches a constant $g_2^* $ for $T \to 0$. The coupling to external spin- and charge density perturbations with wavevector $2k_F$, $h_{\mathrm{SDW}}$ and $h_{\mathrm{CDW}}$, are given by the one-loop corrections
\begin{eqnarray}
\frac{dh_{\mathrm{SDW}}}{ds} &=& \frac{N_0}{2}  \, g^*_{2}  h_{\mathrm{SDW}}\, 2s \, \log \frac{W}{T}\label{hsdweq} \\
\frac{dh_{\mathrm{CDW}}}{ds} &=& \frac{N_0}{2}  \, (g^*_{2} -2 g^*_{1})\,  h_{\mathrm{CDW}} \, 2s \,  \log \frac{W}{T} \label{hcdweq}  \end{eqnarray}
Here we inserted the low temperature fixed-point values for the coupling constants  $g_{1,s}$ and $g_{2,s}$. This is justified for low $T$, since then $g^*_{1}$ and $g^*_{2}$  are approached rapidly for small $s$.
For repulsive $g_1>0$ and $g_2>0$, the SDW coupling grows most strongly when the bare interaction strength $s$ is increased. Integrating Eq. \ref{hsdweq} from $s=0$ to $s=1$, we find the $T \to 0$ power laws for the SDW-coupling and the corresponding susceptibility:
\begin{equation}  h_{\mathrm{SDW}} (T) \propto  T^{-N_0 g^*_2 /2} \, , \qquad \chi_{\mathrm{SDW}} (T) \propto  T^{-N_0 g^*_2} \, \label{plaws} . \end{equation}
These powerlaws are in agreement with results obtained via bosonization \cite{schulz} by expanding the Luttinger liquid parameter $K_\rho$ up to first order in the fixed point coupling $g_2^*$. Likewise, the leading susceptibilities in the $g_1$-$g_2$ plane are reproduced correctly by the IF scheme.

In repulsive Luttinger liquids the presence of a single impurity leads to universal power-laws and scaling functions in the low energy limit. Using a Matsubara frequency cutoff as flow parameter for a spinless fermion model with nearest neighbor interaction and a single static impurity, the computation of spectral properties of single-particle excitations and oscillations in the density profile induced by the impurity or a boundary was shown to provide remarkably accurate results even for moderate coupling strengths in comparison with exact asymptotic and numerical results.\cite{meden}
It turns out that within the IF approach these effects of a single static impurity in a Luttinger liquid are well captured and the respective predictions are confirmed quantitatively. A more detailed comparison is under way.

\section{application to the two-dimensional Hubbard model} \label{implem}

\begin{figure}
\begin{center}
\includegraphics[width=.8\textwidth]{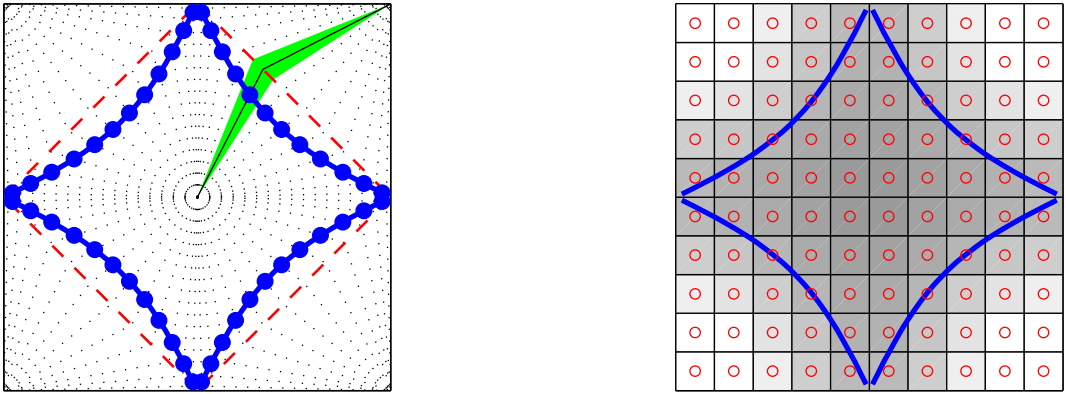}
\end{center}
\caption{Left: $N$-patch discretization for the cutoff and $T$-flow RG schemes. Right plot: Box discretization used for IF. The open circles denote the $N^2$ points for which the coupling function is calculated. The shading indicates the band energy.
 }
\label{discr}
\end{figure}

Here we apply the IF scheme to the two-dimensional $t$-$t'$ Hubbard model on a square lattice and compare the results with those obtained in other RG schemes. The latter have been widely used to identify the leading low-temperature instabilities of the Hubbard model. Practically all numerical 2D implementations  \cite{zanchi,halboth,honsfr,tsai,tflow,katanin} use a patch discretization introduced in this context by Zanchi and Schulz.\cite{zanchi} The coupling function is calculated for two incoming and one outgoing wave vector near or on the Fermi surface, and is taken to be constant within elongated patches that extend roughly perpendicular to the Fermi surface (see Fig.\ \ref{discr}). This approximation is believed to be good, since the leading flow is typically given by the coupling functions connecting particles near the Fermi surface, and because during the flow at least one of the internal lines in the one-loop diagrams is restricted to an increasingly narrow neighborhood of the Fermi surface.  As mentioned in the previous section, the second ''contraction'' feature is absent in the IF, since no cutoff function is used,
and no derivatives of the Fermi function occur in the one-loop diagrams.
Therefore the full BZ contributes to the flow at all values of the flow parameter $g$, and if the couplings near the FS become large it may not be a good approximation to also use them for BZ regions far away from the FS. We are thus forced to use a different discretization of the BZ. This actually \emph{reduces} the degree of approximation caused by the usual projection on the Fermi surface.

We divide the BZ into $N^2$ squares and calculate the flow for wave vectors in the centers of the boxes, keeping the couplings constant within each box. With this prescription we can numerically treat up to $N^2= 18^2= 324$ boxes. Note that this discretization does not correspond to a finite-size system. The internal states in the one-loop diagrams are still taken as those of an infinite lattice and can lie arbitrarily close to the Fermi level.

We focus on band fillings close to van Hove filling and vary the nearest neighbor hopping $t'$.
In the temperature-flow RG the picture is the following  \cite{tflow}: For small $t'$ and near half filling the Fermi surface is well nested, and the flow to strong coupling at low temperatures is mainly of spin-density wave character. For larger $|t'|$ and the FS fixed at the saddle points, $d$-wave pairing tendencies become stronger, while SDW tendencies are suppressed. For $t'<-0.2t$ $d$-wave pairing dominates over SDW. For $t' \to -t/3$ the critical temperature scale for the flow to strong coupling drops by several orders of magnitude. For $t'<-t/3$ it rises again, and the flow to strong coupling is dominated by ferromagnetic tendencies. The temperature-flow RG thus suggests a quantum critical point at $t'\approx -t/3$ between a $d$-wave paired and a ferromagnetic ground state.

\begin{figure}
\begin{center}
\includegraphics[width=.6\textwidth]{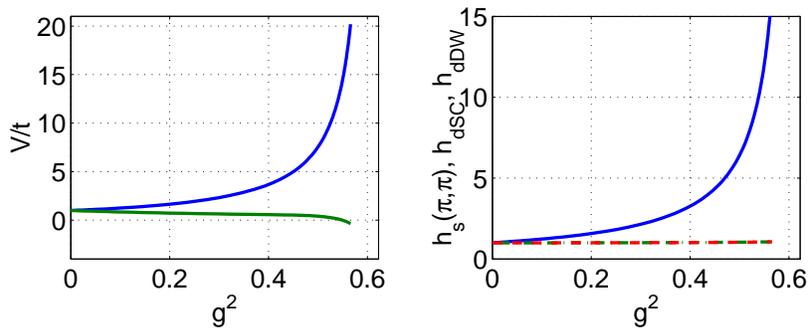}
\end{center}
\caption{Left: Flow of the most attractive and most repulsive coupling constant as a function of the bare interaction strength $g^2=U$ for the perfectly nested case $t'=0$, $\mu=0$. Right: Growth of the coupling to static spin perturbations at $\vec{q}=(\pi,\pi)$ (spin-density wave channel). The two other degenerate lines that rise much slower denote the effective couplings in the $d$-density wave (dDW) and $d$-wave pairing (dSC) channel.
 }
\label{tp0mu0plot}
\end{figure}

\begin{figure}
\begin{center}
\includegraphics[width=.7\textwidth]{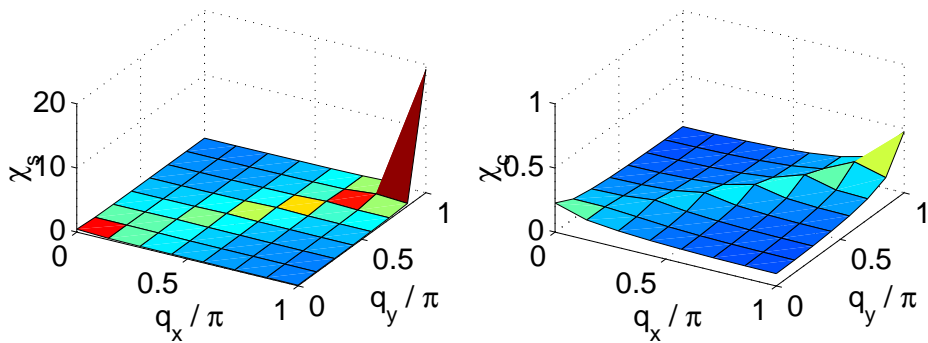}
\end{center}
\caption{Data for half filling and $t'=0$, $T=0.001t$ at the critical bare interaction strength $U=0.57t$ where the couplings diverge. Left: $\vec{q}$-dependence of the interacting spin susceptibility. The peak is at $\vec{q} = (\pi,\pi)$.
Right: $\vec{q}$-dependence of the interacting charge susceptibility. Note the difference in the scale.}
\label{tp0mu0spin}
\end{figure}

At low temperatures, the IF results show (roughly) the same regimes when $|t'|$ is increased, and the FS is fixed at the van Hove points. The results for $t'=0$ are shown in Figs.\ \ref{tp0mu0plot} and \ref{tp0mu0spin}. When the interaction strength is increased for fixed temperature, we find a critical value $g^2_c(T)=U_c(T)$ at which the maximal coupling function diverges.  This value corresponds to the critical interaction strength needed to destroy the weakly coupled state.
For the perfectly nested case, i.e. $t'=0$ and $\mu=0$, the SDW coupling grows strongest when the bare interaction approaches $U_c(T)$.

\begin{figure}
\begin{center}
\includegraphics[width=.7\textwidth]{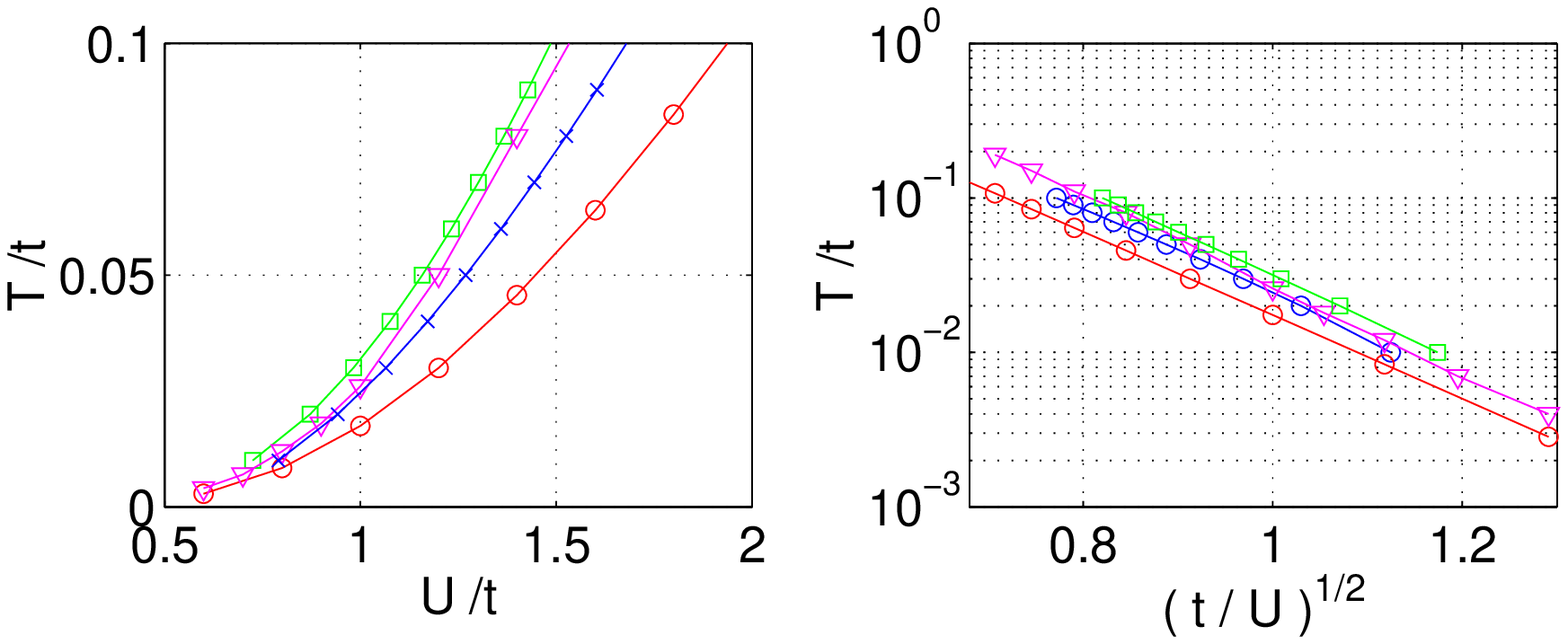}
\end{center}
\caption{Comparison of the characteristic temperatures for the SDW instability
at $t'=0$, $\mu=0$ for several perturbative RG-like schemes. The right plot shows the same data with different $x$-axis. Squares: RPA (360$\times$ 360 lattice in the BZ), circles: $T$-flow (48 patches around the FS), triangles: momentum-shell-RG (1PI-scheme, 48 patches around the FS), crosses: IF for 144 patches in the BZ.}
\label{mu0comp}
\end{figure}

In Fig.\ \ref{mu0comp} we compare the characteristic temperature for the spin-density wave instability  as a function of the bare interaction $U$  for several RG-like schemes with RPA values. In all cases $T_c(U)$ is a monotonically increasing function of $U$, due to its mean-field character. The $T$-flow RG scheme yields the smallest values for $T_c(U)$, RPA the highest. The momentum-shell 1PI scheme lies slightly below the RPA values, and the IF data are roughly in the middle between RPA and $T$-flow.

\begin{figure}
\begin{center}
\includegraphics[width=.6\textwidth]{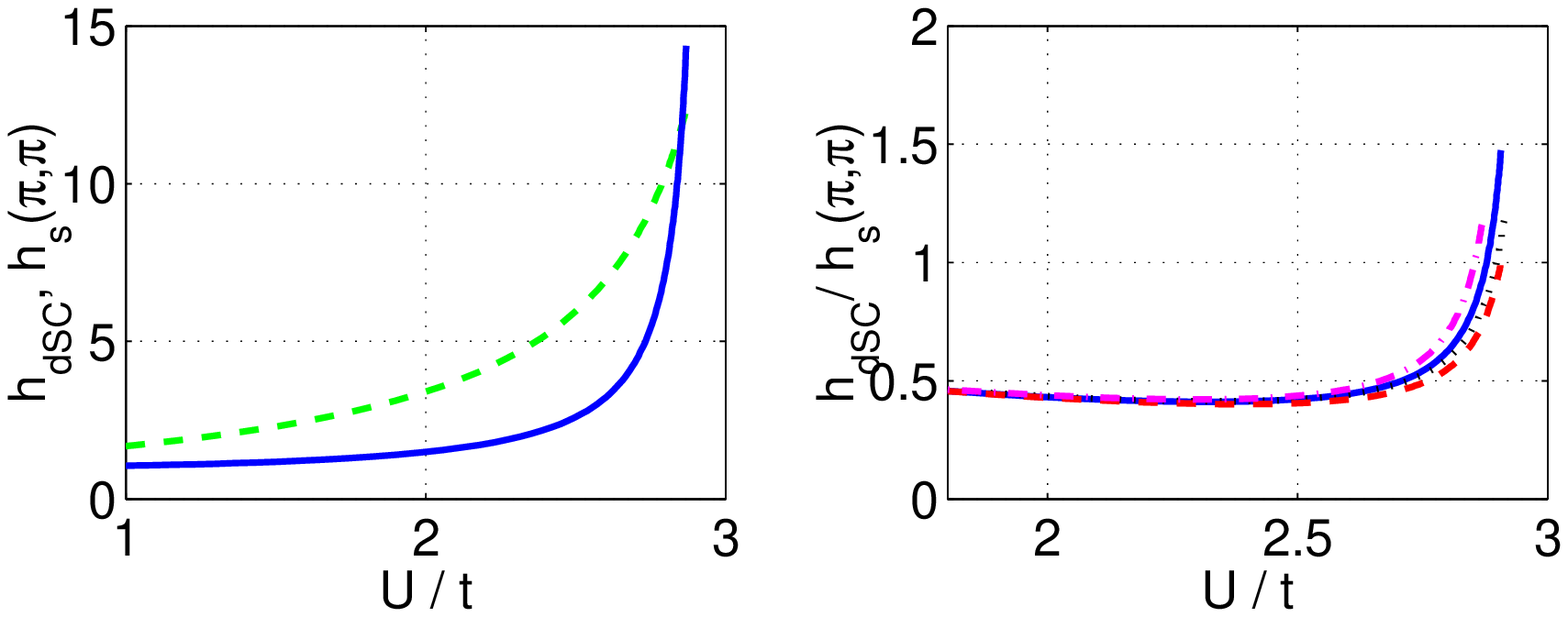}
\end{center}
\caption{Results for the $d$-wave Cooper instability at $\mu=-1.1t$, $t'=-0.3t$ and $T=0.001t$. Left: Flow of effective coupling in $d$-wave pairing (solid line) and AF SDW channel (dashed line) for a 18$\times$18 discretization. Right: Ratio of these two effective coupling constants for different discretizations: solid line 18$\times$18, dashed-dotted line 16$\times$16, dashed line 14$\times$14, dotted line 12$\times$12.}
\label{mu110sizecomp}
\end{figure}

\begin{figure}
\begin{center}
\includegraphics[width=.6\textwidth]{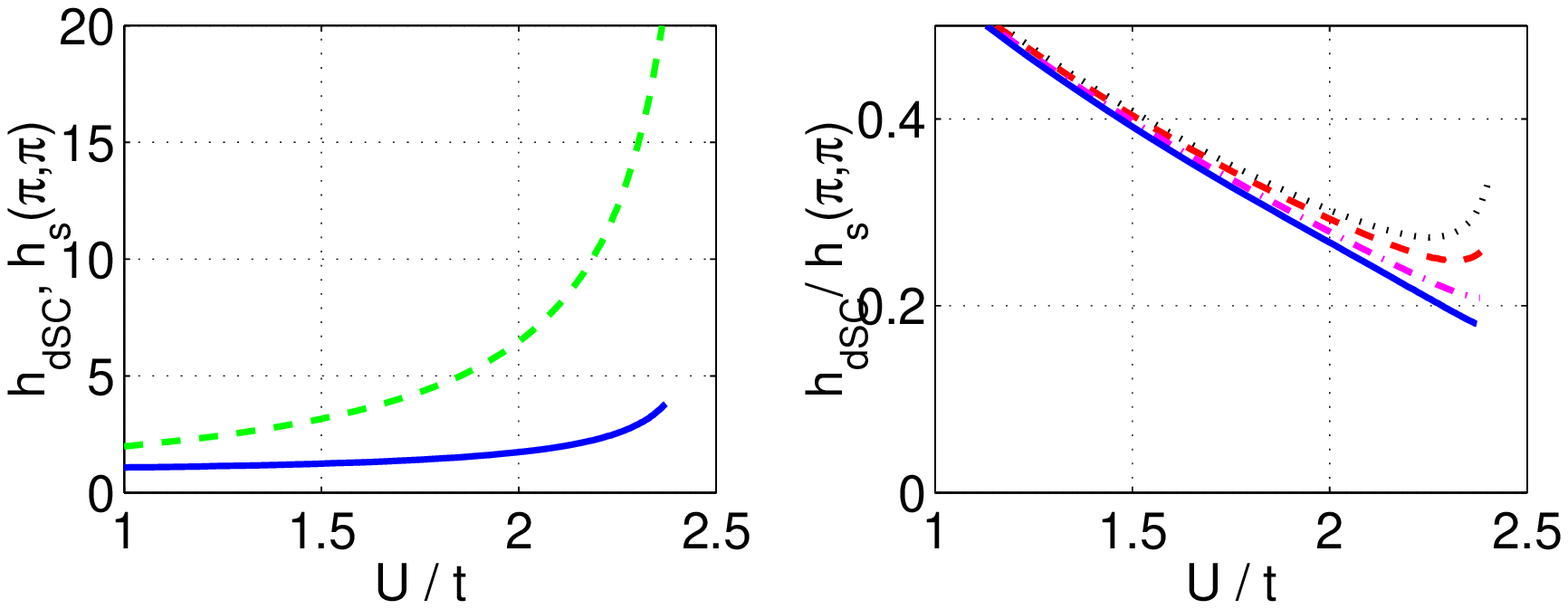}
\end{center}
\caption{The same for van Hove filling at $\mu=-1.2t$, $t'=-0.3t$ and $T=0.001t$. The $(\pi,\pi)$ spin channel may win for large patch numbers.}
\label{mu120sizecomp}
\end{figure}

When we move away from perfect nesting by introducing a next-nearest neighbor hopping amplitude $t'<0$, and change the band filling at low temperature $T=0.001t$, there is an increasing tendency towards $d_{x^2-y^2}$-wave superconducting pairing. For $t'=-0.3t$ this is shown in Figs.\  \ref{mu110sizecomp} and \ref{mu120sizecomp}. In Fig.\ \ref{mu110sizecomp} the band filling is slightly above van Hove filling, and different discretizations show consistently that the effective coupling in the $d$-wave pairing channel grows more strongly than the leading magnetic component at wave vector $\vec{Q} = (\pi,\pi)$.  Fig.\  \ref{mu120sizecomp} shows analogous data at the saddle point filling. With increasing patch number the leading divergence in the SDW channel seems to win over the pairing instability.

\begin{figure}
\begin{center}
\includegraphics[width=.6\textwidth]{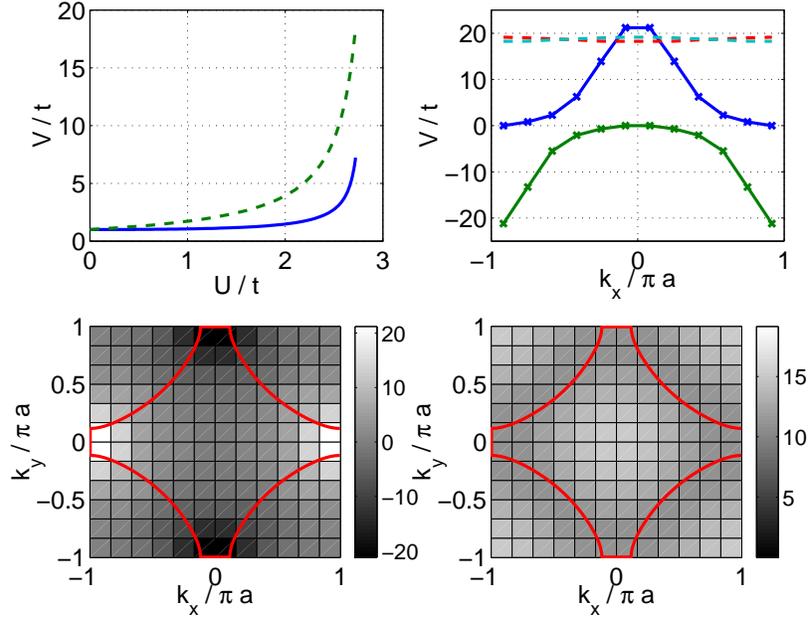}
\end{center}
\caption{IF data for $T=0.001t$ and $\mu=-0.8t$, $t'=-0.25t$ ($\langle n \rangle =0.9$). Upper left plot: Flow of the average coupling to external $d_{x^2-y^2}$-pairing fields, $h_{d\mathrm{-sc}}$ (solid line), and SDW fields, $h^s_{s}(\vec{q}=(\pi,\pi))$ (dashed line), corresponding to the dominant tendencies in this case. Upper right: $d$-wave coupling $h_{d\mathrm{-sc}} (\vec{k},\vec{q}=0)$ (solid lines with crosses) for $\vec{k} = (k_x,0)$ and $\vec{k}=(k_x,\pi)$ at the critical interaction strength. SDW coupling $h^s_{s}(\vec{k}, \vec{q}=(\pi,\pi))$ (dashed lines) for $\vec{k} = (k_x,0)$ and $\vec{k}=(k_x,\pi)$. Lower left/right: $d$-wave and SDW couplings for $\vec{k}$ varying through the full Brillouin zone. The scalebar depicts the strength of the coupling.}
\label{mu80tp25coupl}
\end{figure}

As already mentioned, ignoring lifetime effects possibly makes us overestimate the influence of excitations away from the Fermi surface. In Fig.\ \ref{mu80tp25coupl} we see that the coupling function in the particle-hole channel at wavevector transfer $\vec{q} \approx (\pi,\pi)$, i.e. the SDW channel, is strongly enhanced in all BZ regions, and is largest at the top and bottom of the band. It is likely that in these regions with large band energy the lifetimes of quasiparticles will be short, and the flow without self energy corrections may \emph{overestimate} the SDW tendencies. In contrast to this, the pairing interactions become large only in the vicinity of the Fermi surface. Only a part of this $\vec{k}$-dependence is due to the $d$-wave form factor. 
Due to the discretization scheme we might actually \emph{underestimate} the growth of the pairing correlations, since the coupling function cannot be calculated arbitrarily close to the Fermi surface using only 18\,x\,18 boxes.  Hence, we cannot draw definite conclusions about the leading divergence directly at van Hove filling.
Slightly away from van Hove filling $d$-wave tendencies grow more rapidly than SDW tendencies. This trend is consistently reproduced for all discretizations we used. Keeping in mind that we presumably overestimate the SDW channel and underestimate the pairing tendencies, we find that the dominance of $d$-wave pairing over SDW (as shown in Figs.\ \ref{mu110sizecomp} and \ref{mu80tp25coupl}) is likely to be more robust than suggested by the flow without self energy corrections.\\

\begin{figure}
\begin{center}
\includegraphics[width=.6\textwidth]{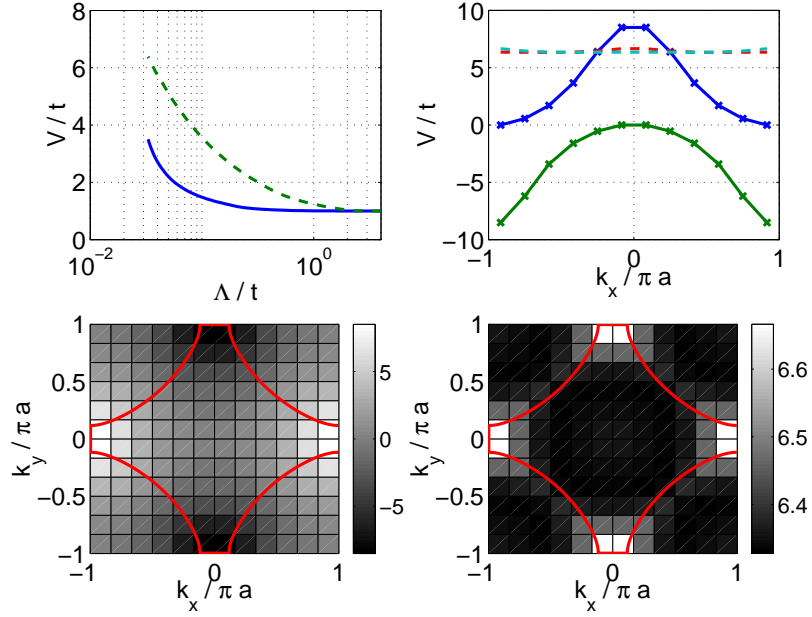}
\end{center}
\caption{Cutoff RG data, again for $T=0.001t$ and $\mu=-0.8t$, $t'=-0.25t$, $U=2.5t$. Upper left plot: Flow vs. infrared cutoff $\Lambda$ of the average coupling to external $d_{x^2-y^2}$-pairing fields, $h_{d\mathrm{-sc}}$ (solid line), and SDW fields, $h^s_{s}(\vec{q}=(\pi,\pi))$ (dashed line), corresponding to the dominant tendencies in this case. Upper right: $d$-wave coupling $h_{d\mathrm{-sc}} (\vec{k},\vec{q}=0)$ (solid lines with crosses) for $\vec{k} = (k_x,0)$ and $\vec{k}=(k_x,\pi)$ at the critical interaction strength . SDW coupling $h^s_{s}(\vec{k}, \vec{q}=(\pi,\pi))$ (dashed lines) for $\vec{k} = (k_x,0)$ and $\vec{k}=(k_x,\pi)$. Lower left/right: $d$-wave and SDW couplings for $\vec{k}$ varying through the full Brillouin zone. The scalebar depicts the strength of the coupling.}
\label{Lmu80tp25coupl}
\end{figure}

In Fig.\ \ref{Lmu80tp25coupl} we show data obtained within the cutoff RG for the one-particle irreducible vertex functions from Ref.\ \onlinecite{honsfr}, using the same discretization scheme as in the IF algorithm. There is good agreement with the IF data in Fig.\ \ref{mu80tp25coupl}. We notice, however, that the SDW channel is a little weaker, and the effective couplings to external SDW fields are now peaked (albeit very mildly) around the FS. Again, this difference occurs because in the cutoff scheme the modes contributing to the flow are restricted to narrow shells around the Fermi surface, meaning that at small scales only those couplings involving modes near the Fermi surface grow most strongly.\\

\begin{figure}
\begin{center}
\includegraphics[width=.6\textwidth]{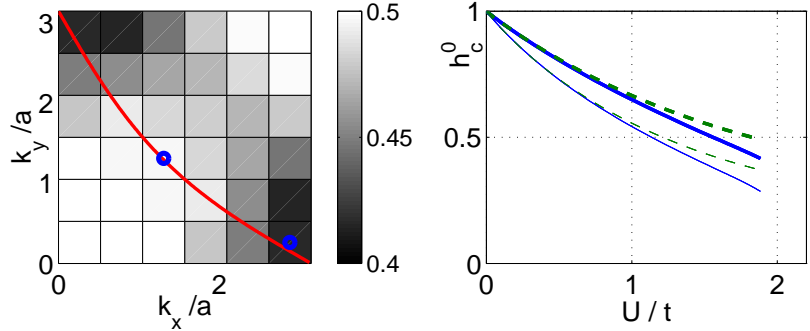}
\end{center}
\caption{IF Data for $T=0.001t$ and van Hove filling $\mu=-t$, $t'=-0.25t$. The left plot shows the coupling to external long-wavelength charge perturbations $h_c (\vec{k},\vec{q} \to 0)$ for $\vec{k}$ varying through a quarter of the Brillouin zone. The tick lines in the right plot shows the flow of  $h_c (\vec{k},\vec{q} \to 0)$ at two points $\vec{k}$. The solid line is for a point close to a saddle point $\vec{k}\approx (\pi, 0)$ and the dashed line for $\vec{k} \approx (1.3,1.3)$ in the BZ diagonal. The thin lines show analogous data for $\vec{q}=(\pi,\pi)$. }
\label{compress}
\end{figure}

Very close to van Hove filling the overall behavior of the IF flow resembles the so-called saddle point regime of Refs. \onlinecite{honsfr} and \onlinecite{furukawa}. The dominant scattering processes near the Fermi surface are those connecting the saddle points, and SDW and $d$-wave pairing susceptibilities both become large. Furthermore, the charge compressibility is suppressed, predominantly in the Brillouin zone regions around $(\pi,0)$ and $(0,\pi)$, similar as in Ref.\ \onlinecite{honsfr}. This can be seen from the flow of the $\vec{k}$-dependent couplings to external static charge perturbations at long wave lengths, $h_c(\vec{k},\vec{q}\to 0)$ (see Fig.\ \ref{compress}). A similar picture is found at half filling. Although at weak coupling these tendencies are rather small, they connect qualitatively well with much stronger effects found by Otsuka et al.  \cite{otsuka} in Quantum Monte Carlo simulations for $U=4t$ and $T=0.2t$.

Upon increasing the temperature in this $d$-wave regime, the instability is pushed to larger bare interactions. For $t'=-0.25t$ and $\langle n \rangle =0.9$ the $d$-wave instability occurs for temperatures below $T=0.0035t$ at $U \le 2.75t$. This is in reasonable agreement with temperature-flow RG results (with a completely different discretization of the Brillouin zone), where the flow diverges at this temperature for $U\approx 2.6t$.
For higher temperatures, the leading divergence appears to occur in the SDW channel at wave vector $(\pi,\pi)$ again, and the interaction needed for the instability barely increases with rising $T$. This indicates that spin fluctuations live on a larger energy scale than pairing tendencies. A similar change in the character of the flow to strong coupling with increasing temperature has been observed in cutoff RG schemes.\cite{honsfr} 
A conservative interpretation suggests the presence of strong spin fluctuations in the normal state above the superconducting ground state (when 3D coupling is present such that long range order is possible). An improved treatment including lifetime effects will give further insights concerning the strength of these effects. A general trend within the present approximations is that in the IF scheme the SDW tendencies seem stronger than in other RG approaches.

\begin{figure}
\begin{center}
\includegraphics[width=.45\textwidth]{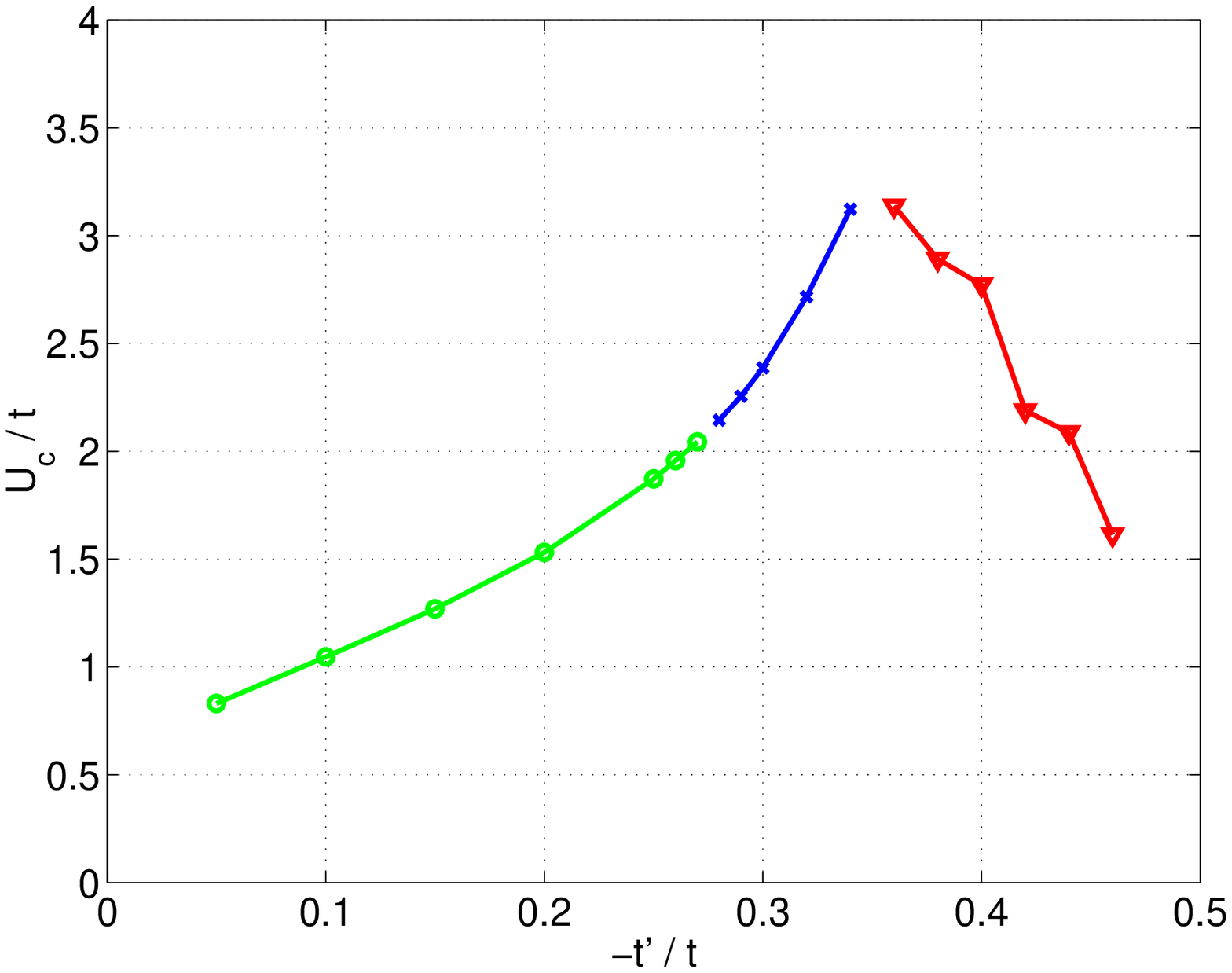}
\end{center}
\caption{Summary of the flow at van Hove filling as a function of $t'$ for $T=0.001t$ and a $12\times 12$ discretization. $U_c$ denotes the critical value of bare interaction strength $U$ at which some components of the coupling function exceed the perturbative range when $U$ is increased from below.
With increasing $|t'|$, for $-t'< 0.2t$ the instability is of AF-SDW type, then it becomes a $d$-wave like Cooper instability. Note that for larger system sizes the boundaries of the SDW-$d$-wave crossover shifts (see discussion in the text). For $-t'> 0.36t$ the dominant instability is ferromagnetic.}
\label{APEvh}
\end{figure}

\begin{figure}
\begin{center}
\includegraphics[width=.45\textwidth]{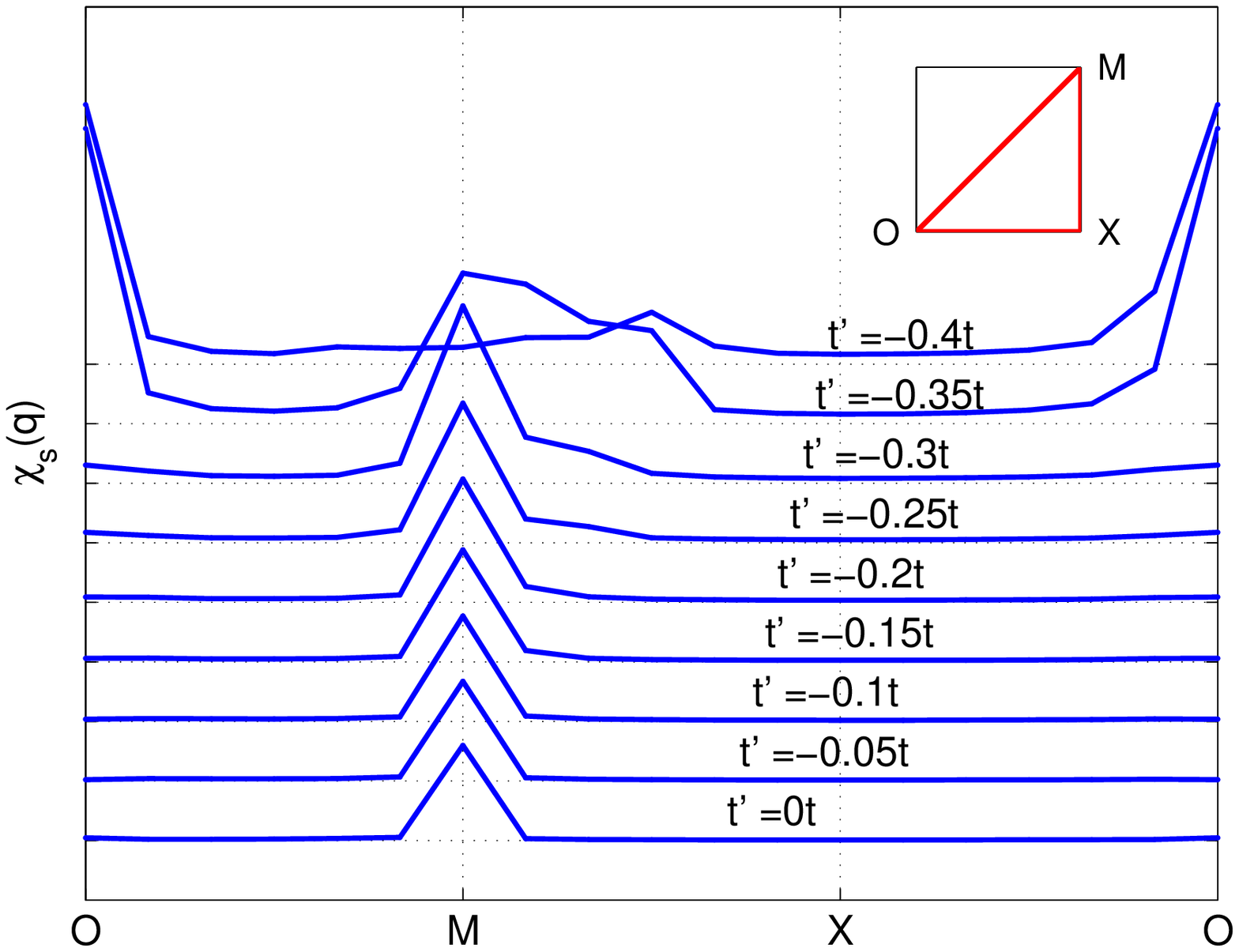}
\end{center}
\caption{Interacting spin susceptibility $\chi_s (\vec{q})$ at van Hove filling  and $T=0.001t$ for different $t'$ and for the  critical interaction strength where the couplings diverge,  using a $12\times 12$ discretization. For $t' > -0.35t$, $\chi_s (\vec{q})$ is peaked at $\vec{q}=(\pi,\pi)=M$ and for $t' \le -0.35t$ it is peaked at $\vec{q}=(0,0)=O$.}
\label{spinsus}
\end{figure}

When $|t'|$ is increased further, the critical interaction strength for the flow to strong coupling increases, and for fixed $U \approx 3t$ the couplings remain bounded down to very low temperatures. For $t' \le -0.36t$ we find that the main instability is in the ferromagnetic channel. The change in the dominant spin correlations can also be seen in Fig.\ \ref{spinsus}, where we plot the spin susceptibility for different $t'$ at van Hove filling along a trajectory in $\vec{q}$-space. The results agree qualitatively well with those of the temperature-flow scheme, which finds the onset of the ferromagnetic regime at $t'\approx -0.33t$ for $U=3t$.

\section{Conclusions and outlook} \label{conclusions}

We have presented a perturbative method to study interacting fermionic systems, the so-called interaction flow (IF) scheme.
Based on the functional renormalization group formalism  \cite{msbook,salmhofer} it sums up all one-loop corrections in an unbiased manner using the bare interaction strength as a continuous flow parameter. We have applied the IF scheme to one-dimensional fermions and the two-dimensional Hubbard model and compared the results with those obtained by cutoff RG and  temperature-flow schemes. There is ample qualitative and in some cases reasonable quantitative agreement, even when very different Brillouin zone discretizations are compared.

In agreement with other RG approaches the IF scheme yields an extended region in parameter space of the 2D $t$-$t'$-Hubbard model
where $d$-wave superconductivity is the dominant low-temperature instability, though it is less prominent, and the tendencies towards SDW ordering appear to be somewhat stronger. This holds in particular for van Hove filling and $t'<-0.2t$, where, opposed to temperature and cutoff flow approaches, the IF scheme does not find a $d$-wave dominated instability. We believe that this difference is due to the fact that in the IF scheme all parts of the BZ contribute to the flow to strong coupling. Neglecting self energy corrections may therefore lead to an overestimate of the SDW channel, whose couplings grow over the whole Brillouin zone. In contrast to this, interactions driving $d$-wave superconductivity become large only in the vicinity of the Fermi surface. This problem does not occur in cutoff RG or temperature-flow schemes. In these approaches only modes in a contracting shell around the Fermi surface contribute in the final stage of the flow.  This way, the IF scheme sheds some light on the inner workings of the previous RG approaches. The inclusion of lifetime effects in the IF scheme will clarify these issues further.

Furthermore, for large absolute values of the nearest neighbor hopping $t'$ the leading instability is in the ferromagnetic channel. This confirms results obtained within the temperature-flow RG and other perturbative approaches.\cite{hankevych}

As new results we have presented detailed wave vector dependences of interactions and susceptibilities over the whole Brillouin zone.

We emphasize that we consider the interaction flow approach an additional tool for the analysis of interacting many-fermion systems. Its concept is as simple as perturbation theory, yet it captures many of the non-trivial effects crucial for the understanding of interacting electron systems. Most importantly the IF scheme includes the coupling between the various one-loop channels. This provides an alternative practicable method to detect and gauge infrared instabilities of fermions in low dimensions. Having several viable techniques at hand may prove useful, as the character of different approximations can be analyzed more clearly.
"True" renormalization group methods like cutoff RG schemes appear to have a broader range of applicability, and seem to be the wider concept. For example, RG methods establish a relationship between systems on different length or energy scales in a quite general way, while the IF flow is by construction restricted to a perturbative regime.

Regarding future applications and extensions of the IF scheme we note the possibility of comparing the results with Quantum Monte Carlo (QMC) data in the 2D Hubbard model and on ladder systems.
Bulut  \cite{bulut} gives detailed results for the irreducible vertex function at $U=4t$ in the two-leg ladder and the 2D Hubbard model,  including wave vector dependence. With the current IF scheme we can easily achieve a comparable $\vec{k}$-space resolution. Hence, it will be interesting to see how the perturbative method compares with QMC. This can give new insights on how the QMC data extrapolate to lower temperatures, that cannot be reached by QMC due to the sign problem.
Moreover, we believe that the inclusion of Fermi surface renormalizations and distortions, as well as lifetime effects are better tractable in the IF scheme, since complications due to a flowing cutoff do not occur. The example of an impurity in a Luttinger liquid described briefly at the end of Sec. \ref{1dexamples} shows that the inclusion of self energy effects in the IF scheme can lead to excellent results. 

All in all, we hope that the IF scheme will open new possibilities in the understanding of interacting many-body systems. \\[1mm]

Acknowledgments: We thank A. Katanin, W. Metzner, and M. Salmhofer for constructive and encouraging discussions.

\end{document}